\documentstyle[aps,prl,twocolumn]{revtex}
\begin{document}
\twocolumn[{
\draft
\widetext
\title{\bf
Scattering~Involving~Prompt~and~Equilibrated~Components,~Information~Theory\\
and~Chaotic~Quantum~Dots}
\author{Harold U. Baranger$^{1}$ and Pier A. Mello$^{2}$}
\address{$^{1}$ AT\&T Bell Laboratories 1D-230, 600
Mountain Avenue, Murray Hill, New Jersey 07974-0636}
\address{$^{2}$ Instituto de F\'{\i}sica, Universidad Nacional
Aut\'{o}noma de M\'{e}xico, 01000 M\'{e}xico D.F., Mexico}
\date{Submitted to Phys. Rev. Lett., February 20, 1995}
\maketitle
\mediumtext
\begin{abstract}
We propose an information-theoretic statistical model to describe the
{\it universal} features of those chaotic scattering processes characterized
by a prompt and an equilibrated component.
The model, introduced in the past in nuclear physics, incorporates
the average value of the scattering matrix to describe the
prompt processes, and satisfies the requirements of
flux conservation, causality, and ergodicity.
We show that the model successfully describes electronic transport
through chaotic quantum dots. The predicted distribution of the
conductance may show a remarkable two-peak structure.
\end{abstract}
}]

\narrowtext

The scattering of waves by complex systems has been a problem
of longstanding interest in physics.

Scattering  of waves by a disordered medium has been studied, for instance,
in optics for a long time \cite{newton&hulst}. Interest in this problem has
been revived, for electromagnetic waves and for electrons, in relation with
the phenomenon of localization, with a host of exciting new
features \cite{sheng,altshuler(book)}. Here, the diffusion time throughout
the medium is the single important characteristic time.

Examples of quantum-mechanical scattering by complex systems can also
be found in nuclear and molecular physics. It is amazing that one can
describe the scattering of a nucleon by an atomic nucleus---a complicated
many-body problem---in terms of {\it two distinct time scales}:
a {\it prompt} response arising
from direct processes and a time-delayed one arising from the formation of
an {\it equilibrated} compound nucleus.
The prompt response is slowly varying in energy and is described by the
energy averaged, or {\it optical}, scattering amplitudes; the equilibrated
response is the difference from this energy average and is amenable to a
statistical analysis \cite{feshbach&ericson}.
Further examples of physical processes of this type, and studied in terms
of similar notions, can be found in molecular physics, with interesting
applications to chemistry \cite{levine&miller}.

Most remarkably, features similar to those appearing in scattering from
nuclei also occur in
quantum-mechanical scattering from simple one-particle systems
\cite{Gutz83,Blu88}. An example is a particle
scattering from a cavity of dimensions larger than the
wavelength, in which the classical dynamics is chaotic.
One experimental realization of such systems are the ballistic quantum
dots \cite{ReviewMes,JalBarSto,expt}, microstructures in which both the
phase-coherence length and the elastic mean free path exceed the system
dimensions; the dot acts as a
resonant cavity and the leads as electron waveguides.
Previous work on these systems has implicitly assumed the absence of any
prompt response.

Our purpose is to propose a model describing the {\it universal} features
that appear in any chaotic scattering process involving a prompt and an
equilibrated component.
This model was introduced  in the past in the context of nuclear physics,
using an {\it information-theoretic} approach
\cite{mello-pereyra-seligman} based on the mathematical development in
Ref. \cite{hua}. Here we show that the same theoretical framework is
successful in the description of electronic transport through ballistic
quantum dots. We build on previous work which used this model in describing
chaotic scattering \cite{DorSmi} and in simulating phase-breaking in
quantum dots \cite{PietCarlo2}.

Semiclassical \cite{Blu88,JalBarSto}, field-theoretic \cite{fieldtheory},
and  random-matrix \cite{fieldtheory,harold-pier,rodolfoEPL94} approaches
have been used to describe quantum transport through ballistic quantum dots.
In Refs. \cite{harold-pier,rodolfoEPL94} the statistics of the
problem was described by assigning to the quantum scattering matrix $S$
an ``equal a priori distribution'',  consistent
with the symmetry requirements. This ``invariant measure"
\cite{ReviewRMT} defines the ``circular ensemble" of $S$ matrices.
The results for the {\it ensemble}
average, variance and probability density of the conductance
were found, in Ref. \cite{harold-pier}, to agree with
a statistical analysis of the numerically obtained conductance
of a chaotic cavity connected to two waveguides, sampled along the {\it energy}
axis. It was assumed
that  one could neglect ``direct processes'' caused by short
trajectories that would give a nonvanishing energy averaged, or optical,
$S$-matrix; this was enforced in the simulations with
two stoppers which block direct transmission between the
leads and whispering-gallery trajectories. Although the general
problem may contain a range of relevant ``time delays'', we  present below
an improvement on the above model in terms of
{\it two very different time scales}, in a vein similar to the nuclear
scattering problem above. We first summarize the
information-theoretic approach \cite{mello-pereyra-seligman}.

{\it Information-Theoretic Approach}---
A quantum scattering problem is described by its $S$ matrix,
which, for scattering involving two leads, each with width $W$ and
$N$ transverse modes or channels, is $n=2N$-dimensional and has the structure
\begin{equation}
\label{S}S=\left( \begin{array} {cc}
r & t^{\prime}\\
t & r^{\prime}
\end{array} \right) \; .
\end{equation}
Here, $r$, $t$ are the $N \times N $ reflection and transmission
matrices for incidence from the left and $r^{\prime}$,
$t^{\prime}$  from the right.
Current conservation requires $S$ to be unitary,
$SS^{\dagger} =I$; for time-reversal symmetry (as is realized
in the absence of a magnetic field) and no spin, $S$ is symmetric.

Our starting point is $d \mu^{(\beta )} (S)$, the {\it invariant} measure
under the symmetry operation for the universality class $\beta$ in question.
The operation is $S^{\prime }=U_0 S V_0 $, where $U_0 $, $V_0 $ are
arbitrary fixed unitary matrices in the case of unitary
$S$ matrices [the circular unitary ensemble ($\beta =2$)], with the
restriction $V_0 = U_0 ^T $ in the case of unitary symmetric $S$ matrices
[the circular orthogonal ensemble ($\beta =1$)]. $d \mu^{(\beta)}(S)$ can be
written explicitly in several different
representations \cite{hua,harold-pier,rodolfoEPL94}.

The ensemble average of $S$, and hence the prompt component, vanishes when
evaluated with the invariant measure.  Ensembles in which $\langle S \rangle $
is nonzero contain more {\it information} than the circular ensembles;
they are constructed by multiplying the invariant measure by a function of $S$
to give the differential probability
\begin{equation}
\label{dP(S)}d P_{\langle S \rangle }^{(\beta )}
= p_{\langle S \rangle }^{(\beta)}(S) \; d \mu^{(\beta )}(S) \; .
\end{equation}
The information ${\cal I}$ associated with the above
probability distribution is defined as \cite{khinchin}
\begin{equation}
\label{information}
{\cal I}[p_{\langle S \rangle }]
\equiv \int p_{\langle S \rangle }(S) \;
\ln [ p_{\langle S \rangle }(S) ] \; d \mu (S) \; .
%{\cal I}[p_{\langle S \rangle }^{(\beta )}]
%\equiv \int p_{\langle S \rangle }^{(\beta )}(S) \;
%\ln [ p_{\langle S \rangle }^{(\beta )}(S) ] \;  d \mu (S) \; .
\end{equation}

Far from channel thresholds, the $S$-matrix is {\it analytic} in the
upper half of the complex-energy-plane (causality). We also require that the
ensemble be {\it ergodic} \cite{yaglom}, so that energy averages can be
replaced by ensemble averages.  These {\it analyticity-ergodicity
requirements (AE)} imply the {\it reproducing property}
\begin{equation}
\label{ReproducingProperty}
f(\langle S \rangle)=\int f(S) d P_{\langle S \rangle}(S)
\; \; ,
\end{equation}
for a function $f(S)$ analytic in its argument (expandable in  a power series
in $S$ not involving $S^{\ast }$).
The probability density known as {\it Poisson's kernel},
\begin{equation}
\label{poisson}p_{\langle S \rangle }^{(\beta)}(S) = V_{\beta }^{-1}
\frac {[{\rm det }
(I-\langle S \rangle \langle S^{\dagger} \rangle ) ]
^{(\beta n +2-\beta )/2}}
{\mid {\rm det }
(I-S \langle S^{\dagger } \rangle )\mid ^{\beta n +2 - \beta}}
\end{equation}
where $V_{\beta}$  is a normalization constant, satisfies the reproducing
property Eq. (\ref{ReproducingProperty}) \cite{hua},
and the associated information is less than or equal to that
of any other probability density satisfying the AE requirements for the
same $\langle S \rangle $ \cite{mello-pereyra-seligman}.
Recently, this probability density for the $S$-matrix has been derived
from a statistical distribution for the Hamiltonian \cite{piet},
explaining the coincidence noticed between the two approaches
\cite{mello-pereyra-seligman}.
{\it Thus, Poisson's kernel describes those physical situations
in which (a) the details are irrelevant except for the average $S$-matrix
and (b) the requirements of flux conservation, time-reversal
invariance (when applicable), and AE must be met.}

For $n=1$, when the system is a cavity
connected to the outside by only a one-mode lead and
$S={\rm e}^{i \theta }$ describes reflection back into the
same lead, the ensemble is {\it uniquely} determined by AE
and a specified value of $\langle S \rangle $ and is given by
Poisson's kernel \cite{reyes}. For $n>1$ the additional
minimum-information criterion explained above is needed to determine
the ensemble.

{\it Transport through Quantum Dots---}
In terms of the $S$ matrix, the conductance for spinless particles is
\cite{ReviewMes}
\begin{equation}
\label{G}G=(e^2 /h)T = (e^2 /h) {\rm Tr}[tt^{\dagger }].
\end{equation}
Thus, in applying Poisson's kernel Eq. (\ref{poisson}) to electronic transport
through quantum dots, we need the probability distribution $w(T)$
of the total transmission $T$; i.e.
\begin{equation}
w(T) = \int \delta (T- {\rm Tr}[tt^{\dagger }] )
p_{\langle S \rangle }^{(\beta)}(S) \; d \mu^{(\beta )}(S) \; \; .
\label{w(T)}
\end{equation}
We study below the case $n=2$, for which there is only one mode in each lead.

We first discuss $\beta =2$. Denote the elements of $\langle S \rangle$ by
\begin{equation}
\label{Sbarelements}
\langle S \rangle = \left(
\begin{array}{cc}
         x & w \\
         z & y
\end{array}
\right)
\end{equation}
where $w$, $x$, $y$, $z$ are complex numbers with $X \equiv |x| \leq 1$, etc.
If there is no prompt transmission,
$w=z=0$, we can perform the integrations analytically, yielding
\begin{eqnarray}
\label{w(T)beta2Sdiag}w&&(T)=(1-X^2)^2 (1-Y^2)^2\nonumber\\
&&\times \lbrace (1-X^4Y^4)(1-X^2Y^2) \nonumber \\
&&\phantom{.......}-(1-T)[(X^2 +Y^2)(1-6X^2Y^2 + X^4Y^4)\nonumber\\
&&\phantom{......................} +4X^2Y^2(1+X^2Y^2)] \nonumber \\
&&\phantom{.......}+(1-T)^2[ (1+X^2Y^2)(6X^2Y^2 -X^4 -Y^4)\nonumber\\
&&\phantom{......................} -4X^2Y^2(X^2 +Y^2)]  \nonumber\\
&&\phantom{.......}+(1-T)^3(X^2 +Y^2)(X^2-Y^2)^2 \rbrace \nonumber \\
&&\times \lbrace (1-X^2Y^2)^2 -2(1-T)[(1+X^2Y^2)(X^2 +Y^2)\nonumber\\
&&\phantom{......................}-4X^2Y^2] \nonumber \\
&&\phantom{.......}+(1-T)^2(X^2-Y^2)^2 \rbrace ^{-5/2}.
\end{eqnarray}
For $x=y$  the above result reduces to
\begin{eqnarray}
\label{w(T)beta2Sdiagx=y}w&&(T)=(1-X^2)\nonumber\\
&&\times\frac
{(1-X^4)^2 +2X^2(1+X^4)T +4X^4T^2}
{[(1-X^2)^2+4X^2T]^{5/2}} \; .
\end{eqnarray}
Calling $\Gamma =1-X^2$, Eq. (\ref{w(T)beta2Sdiagx=y}) gives, in the
limit $\Gamma <<1$, the result of Ref. \cite{PietCarlo1}.
In the opposite case of no prompt component to the reflection, $x=y=0$,
$w(T)$ is related to that of
Eq. (\ref{w(T)beta2Sdiag}) by the replacements
$x\rightarrow w$, $y\rightarrow z$, $T \rightarrow 1-T$.
%
%\begin{equation}
%\label{ReplacementForSoffdiag}
%x\rightarrow w \; , \; \;
%y\rightarrow z \; , \; \;  T \rightarrow 1-T \; .
%\end{equation}
%
For nonzero $x$, $y$, $w$, $z$, we can express the result in terms of a
single angular
integration; it will not be given here because of lack of space.

We now discuss the case $\beta  =1$. When $\langle S \rangle$  is diagonal
and $y=0$, one finds (see Eq. (5.13) of Ref. \cite{mello-pereyra-seligman})
%\begin{eqnarray}
%\label{w(T)beta1Sdiagy0}w(&&T)=(1/2)(1-X^2)^{3/2}\nonumber\\
%&&\times T ^{-1/2}\; F(\frac{3}{2},\frac{3}{2};1;(1-T)X^2)
%\end{eqnarray}
\begin{equation}
\label{w(T)beta1Sdiagy0}
w(T)= \frac{(1-X^2)^{3/2}}{2 \sqrt{T}}
F(\frac{3}{2},\frac{3}{2};1;(1-T)X^2)
\end{equation}
where $F$ is a hypergeometric function. When
$\langle S \rangle$ is diagonal and $x=y$, we find
%\begin{eqnarray}
%\label{w(T)beta1Sdiagx=y}w(&&T)=C T^{-1/2}\nonumber\\
%&&\times \Biggl\langle \frac {F(\frac{3}{2},\frac{3}{2};1;|E|^2)}
%{[1+(1-T)X^2-2\sqrt{1-T}X\cos \psi]^{3/2}}\Biggr\rangle _{\psi }\; ,
%\end{eqnarray}
\begin{equation}
\label{w(T)beta1Sdiagx=y}
w(T)= \frac{C}{\sqrt{T}}
\Biggl\langle \frac {F(\frac{3}{2},\frac{3}{2};1;E^2)}
{[1+(1-T)X^2-2\sqrt{1-T}X\cos \psi]^{3/2}}\Biggl\rangle _{\psi} ,
\end{equation}
where $\langle \rangle _{\psi}$ indicates an average over
$\psi \in [0, 2\pi ]$ and
%\begin{equation}
%\label{E^2}
%E^2=X^2\Biggl\lbrack 1-\frac {(1-X^2)T}
%{1+(1-T)X^2-2\sqrt{1-T}X\cos \psi }\Biggr\rbrack \; .
%\end{equation}
\begin{equation}
\label{E^2}
E^2 = X^2 \frac
{1 - T + X^2 - 2\sqrt{1-T}X\cos \psi }
{1 - TX^2  + X^2 - 2\sqrt{1-T}X\cos \psi } .
\end{equation}
When $\langle S \rangle$ is offdiagonal ($x=y=0$) with $w=z$, we find
\begin{eqnarray}
\label{w(T)Soffdiag}w(T)&&=(1/2)(1-Z^2)^3 T^{-1/2}
\Bigl\langle (1+2(2T-1)Z^2+Z^4 \nonumber \\
&& -4\sqrt{T}Z(1+Z^2)\cos \psi
+4Z^2\cos ^2 \psi)^{-3/2}\Bigr\rangle _{\psi }.
\end{eqnarray}

For several cases, these complicated distributions are plotted
in Fig. 1 and will be discussed in connection with the numerical results
below.

{\it Numerical Results---}
We have computed the conductance for several stadium billiards,
sketched in Fig. 1, using the methods of Ref. \cite{numerical}.
$w(T)$ was found by sampling in an energy window much
larger than the energy correlation length but smaller
than the interval over which the prompt response changes (so that
``stationarity", a condition for ergodicity \cite{yaglom}, is attained) and by
using several slightly different structures. Typically we used 200 energies
in $kW/\pi \in [1.6,1.8]$ and 10 structures found by changing
the height or angle of the convex ``bumper'' in Fig. 1.
Thus we rely on ergodicity to compare the numerical distributions to the
ensemble averages of random-matrix theory. In each case the optical
$S$-matrix was extracted directly from the numerical data; in this sense the
theoretical curves shown below are parameter free.

We first consider a simple half-stadium with collinear leads
at low magnetic field
($BA/ \phi_0 = 2$, $r_c = 55\; W$, where $A$ is the area of the cavity,
$r_c$ is the cyclotron radius, and $W$ is the width of the leads):
$w(T)$ is nearly uniform [Fig. 1(a)],
and $\langle S \rangle$ is small because direct trajectories are negligible
in this large structure.  We thus obtain good agreement with the circular
unitary ensemble prediction, as in previous work \cite{harold-pier}.

In order to increase $\langle S \rangle$ we modify the situation in three
ways: (1) introduce potential barriers at the openings of the leads into the
cavity (dashed lines in structures of Fig. 1), (2) increase the magnetic
field, and (3) extend the leads into the cavity.
The barriers (chosen so that the transmission of each barrier is $1/2$)
increase the direct reflection and thus skew the distribution towards
small $T$ [Fig. 1(b)].
The large magnetic field ($BA/ \phi_0 = 80$, $r_c = 1.4 \: W$)
increases one component of the direct transmission--- the one corresponding
to skipping orbits along the lower edge--- and thus skews the distribution
towards large $T$ [Fig. 1(c)].
Finally, extending the leads into the cavity increases the direct transmission
in both directions and thus also skews the distribution towards large $T$
[Fig. 1(d)].
Note the excellent agreement with the information-theoretic model.
In panels (b)-(d) the curve plotted is the analytic expression of
Eq. (\ref{w(T)beta2Sdiag}) and the corresponding one for direct transmission.

By combining several of these modifications, different $\langle S \rangle$
and so different distributions can be produced. First,
by using extended leads with barriers at their ends, one can cause
both prompt transmission and reflection: this case, Fig. 1(e), is in good
agreement with the prediction of the full Poisson's kernel. Finally,
increasing the magnetic field in this structure produces a large average
transmission and a large average reflection.
The resulting $w(T)$, Fig. 1(f), has a surprising two-peak structure: one peak
near $T=1$ caused by the large direct transmission and another near $T=1/2$.
Even in this unusual case, the prediction of the information-theoretic model
is in excellent agreement with the numerical result.
In these last two cases (e,f) the analysis was performed independently
over four intervals
of 50 energies each (since the four intervals show slightly different
$\langle S \rangle$'s ) and the four sets of theoretical and numerical data
were then superimposed.

{\it Discussion---}
In addition to the structures shown in Fig. 1 we have studied cavities whose
level density is not large enough for stationarity, and hence ergodicity, to
hold. In this case, a sample taken across independent cavities for a
{\it fixed energy} shows excellent agreement with Poisson's kernel.  This
suggests that
the reproducing property Eq. (\ref{ReproducingProperty}) may be valid even
in the absence of ergodicity; the reason for this is not understood.

In Ref. \cite{harold-pier} we found that increasing the
magnetic flux through the structure beyond a few flux quanta spoiled the
agreement with the circular ensemble; we now know that a nonzero
$\langle S \rangle$ is generated and that the present model describes the
data very well. The excellent agreement found here with a flux as high as
80 suggests extending the analysis to the quantum Hall regime.

We close by noting that the above $w(T)$'s  should be experimentally
accessible in structures where phase-breaking is small enough.
Experimentally one can sample the conductance distribution by
varying the energy or shape of the structure with an external gate voltage
\cite{Chan95},
much as we did in collecting the numerical results. The barrier at the
opening of the leads can be realized by designing a pincher gate and,
of course, obtaining a sufficiently high magnetic field is standard.

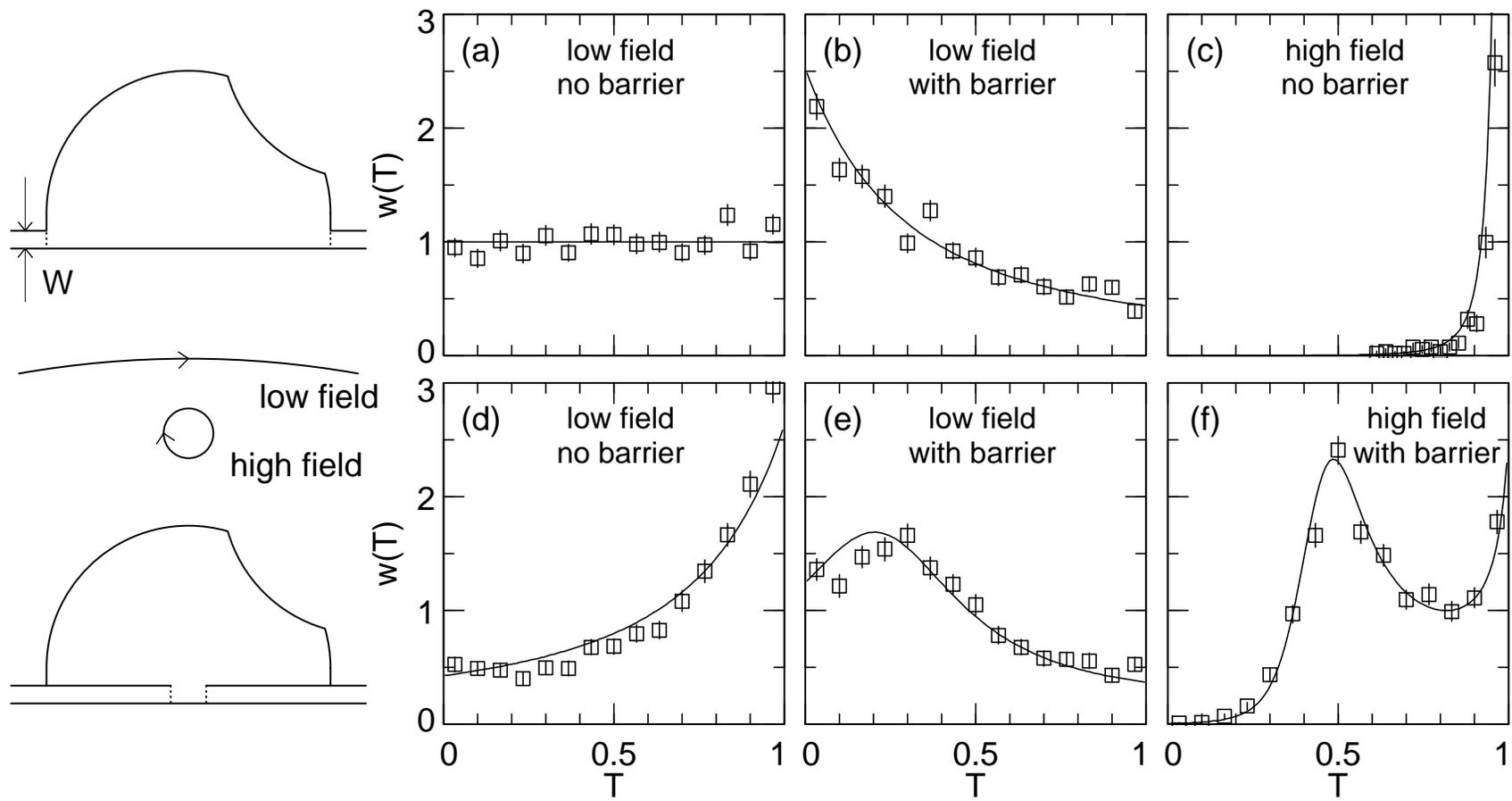
\begin{figure}
\caption{The distribution of the transmission coefficient for $N=1$ in
a simple half-stadium (top row) and a half-stadium with leads extended
into the cavity (bottom row). The magnitude of the magnetic field and
the presence or absence of a potential barrier at the entrance to the leads
(marked by dotted lines in the sketches of the structures) are noted in each
panel. Cyclotron orbits for both fields, drawn to scale, are shown on the left.
The squares with statistical error bars are the numerical
results; the lines are the predictions of the information-theoretic model,
parametrized by an optical $S$-matrix extracted from the numerical data.
The agreement is good in all cases.}
\end{figure}


\begin{references}

\bibitem{newton&hulst}
R. G. Newton,
{\it Scattering Theory of Waves and Particles} (McGraw-Hill, New York, 1966);
H. C. van de Hulst,
{\it Light Scattering from Small Particles} (Dover, New York, 1981).

\bibitem{sheng}
P. Sheng, {\it Scattering and Localization of Classical Waves in
Random Media} (World Scientific, Singapore, 1990).

\bibitem{altshuler(book)}
B. L. Altshuler, P. A. Lee and R. A. Webb, eds., {\it Mesoscopic
Phenomena in Solids} (North-Holland, Amsterdam, 1991).

\bibitem{feshbach&ericson}
H. Feshbach, C. E. Porter, and  V. F. Weisskopf,
Phys. Rev. {\bf 96}, 448 (1954);
T. Ericson, Ann. of Phys. (N.Y.) {\bf 23}, 390 (1963);
H. Feshbach, {\it Topics in the Theory of Nuclear Reactions},
in {\it Reaction Dynamics} (Gordon and Breach, New York, 1973).

\bibitem{levine&miller}R. D. Levine, {\it Quantum Mechanics of
Molecular Rate Processes} (Oxford University Press, Oxford, 1969), Ch. 3.5;
W. H. Miller, J. Chem. Phys. {\bf 65}, 2216 (1976).

\bibitem{Gutz83}
M. C. Gutzwiller, Physica D {\bf 7}, 341 (1983).

\bibitem{Blu88}
 R. Bl\"umel and U. Smilansky,
Phys. Rev. Lett.  {\bf 60}, 477 (1988); {\bf 64}, 241 (1990);
Physica D {\bf 36}, 111 (1989).

\bibitem{ReviewMes}
For a review see
C. W. J. Beenakker and H. van Houten in
{\it Solid State Physics}, edited by H. Ehrenreich and D. Turnbull
(Academic, New York, 1991), Vol. 44, pp. 1-228.

\bibitem{JalBarSto}
R. A. Jalabert, H. U. Baranger, and A. D. Stone,
Phys.  Rev. Lett. {\bf 65}, 2442 (1990);
H. U. Baranger, R. A. Jalabert, and A. D. Stone,
Phys. Rev. Lett. {\bf 70}, 3876 (1993); Chaos {\bf 3}, 665 (1993).

\bibitem{expt}
Representative experiments are reported in
\newline C. M. Marcus, {\it et al.}, Phys. Rev. Lett. {\bf 69}, 506 (1992);
\newline M. W. Keller, {\it et al.}, Surf. Sci. {\bf 305}, 501 (1994);
\newline A. M. Chang, {\it et al.}, Phys. Rev. Lett. {\bf 73}, 2111 (1994);
\newline M. J. Berry, {\it et al.}, Phys. Rev. B {\bf 50}, 17721 (1994).

\bibitem{mello-pereyra-seligman}
P. A. Mello, P. Pereyra, and T. H. Seligman,
Ann. Phys. (N.Y.) {\bf 161}, 254 (1985);
W. A. Friedman and P. A. Mello,
Ann. Phys. (N.Y.) {\bf 161}, 276 (1985).

\bibitem{hua}
L. K. Hua, {\it Harmonic Analysis of Functions of Several Complex Variables
in the Classical Domain} (Amer. Math. Soc., Providence RI, 1963).

\bibitem{DorSmi}
E. Doron and U. Smilansky,
Nucl. Phys. A {\bf 545}, C455 (1992).

\bibitem{PietCarlo2}
P.W.Brouwer and C.W.J.Beenakker, preprint (10/94).

\bibitem{fieldtheory}
A. Pluhar, H. A. Weidenm\"uller, and J. A. Zuk,
Phys. Rev. Lett. {\bf 73}, 2115 (1994);
K. Efetov, preprint (July, 1994);
V. N. Prigodin, K. B. Efetov, and S. Iida,
Phys. Rev. Lett. {\bf 71}, 1230 (1993);
S. Iida, H. A. Weidenm\"uller, and J. A. Zuk,
Annals of Phys. {\bf 200}, 219 (1990).

\bibitem{harold-pier}
H. U. Baranger and P. A. Mello,
Phys. Rev. Lett. {\bf 73}, 142 (1994);
Phys. Rev. B, February 15, 1995.

\bibitem{rodolfoEPL94}
R. A. Jalabert, J.-L. Pichard, and C. W. J. Beenakker,
Europhys. Lett. {\bf 27}, 255 (1994);
R. A. Jalabert and J.-L. Pichard, preprint (September 1994).

\bibitem{ReviewRMT}
M. L. Mehta, {\it Random Matrices} (Academic, New York, 1991);
C. E. Porter, {\it Statistical Theories of Spectral Fluctuations}
(Academic, New York, 1965).

\bibitem{khinchin}
A. I. Khinchin, {\it Mathematical Foundations of Information Theory}
(Dover, New York, 1957).

\bibitem{yaglom}
A. M. Yaglom, {\it An Introduction to the Theory of
Stationary Random Functions} (Prentice-Hall, New York, 1962).

\bibitem{piet}
P. W. Brouwer, preprint (January, 1995).

\bibitem{reyes}
J. de los Reyes, P. A. Mello and T. H. Seligman,
Z. Physik A {\bf 295}, 247 (1980).

\bibitem{PietCarlo1}
P.W.Brouwer and C.W.J.Beenakker, preprint (6/94).

\bibitem{numerical}
H. U. Baranger, D. P. DiVincenzo, R. A. Jalabert, and A. D. Stone,
Phys. Rev. B {\bf 44}, 10637 (1991). For the results here, we
used $W/a = 10$ and $ka \approx 0.55$.

\bibitem{Chan95}
I. H. Chan, {\it et al.}, Phys. Rev. Lett. (in press).

\end{references}
\end{document}